\documentclass[prd,reprint,amsmath,nofootinbib]{revtex4-1}
\pdfoutput=1

\usepackage{ifthen}
\usepackage{epsfig}
\usepackage{booktabs} 
\usepackage{natbib}
\usepackage{wasysym}
\usepackage{feynmp}
\usepackage{slashed} 

\usepackage{color}

\newcommand{\beqra}{\begin{eqnarray}}
\newcommand{\eeqra}{\end{eqnarray}}
\newcommand{\beq}{\begin{equation}}
\newcommand{\eeq}{\end{equation}}

\begin{document}

\title{New Spectral Features from Bound Dark Matter}

\author{Riccardo Catena}
\email{catena@chalmers.se}
\affiliation{Chalmers University of Technology, Department of Physics, SE-412 96 G\"oteborg, Sweden}
\author{Chris Kouvaris}
\email{kouvaris@cp3.sdu.dk}
\affiliation{CP$^3$-Origins, University of Southern Denmark, Campusvej 55, DK-5230 Odense, Denmark}

\begin{abstract}
We demonstrate that dark matter particles gravitationally bound to the Earth  can induce a characteristic nuclear recoil signal at low energies in direct detection experiments.~The new spectral feature  we predict can provide the ultimate smoking gun for dark matter discovery for experiments with positive signal but unclear background.~The new feature is universal, in that the ratio of bound over halo dark matter event rates at detectors is independent of the dark matter-nucleon cross section. 
\\[.1cm]
{\footnotesize  \it Preprint: CP3-Origins-2016-003 DNRF90}
\end{abstract}


\maketitle

\section{Introduction}
Milky Way Dark Matter (DM) particles are expected to scatter against nuclei in underground detectors while crossing the Earth.~The direct detection of DM collisions with nuclei in Earth based detectors is the goal of a plethora of experiments in operation today.~Although simple in principle, direct detection is a difficult task in practice. It is challenging to distinguish the potential DM signal from background events. It will become even more challenging in the near future since the exposure of the experiments will soon be sufficient to probe solar and atmospheric neutrinos which could fake the DM signal. These challenges require the develpment of new detection strategies that could identify the DM signal beyond doubt.

In this paper we show that DM particles gravitationally bound to the Earth create a unique feature in the low energy spectrum in direct detection experiments that could be the ``smoking gun" of DM discovery. In particular, DM particles that cross the surface of the Earth can get captured due to underground collisions and become gravitationally bound to the Earth. After their capture, these particles travel on elliptical orbits that cross periodically the surface of the Earth. Particles in orbits  that travel only small distances underground, can remain in these orbits for a very long period before they scatter again. Therefore a large amount of DM particles can accumulate in these particular orbits during the lifetime of the Earth. These particles can create a significant distortion of the  recoil energy spectrum in direct detection experiments. This new spectral feature can provide the ultimate verification of DM discovery. It can also help probe DM below the neutrino floor that direct DM experiments will soon face. The presence of this new spectral feature is independent of the DM-nucleon cross section. Gravitationally bound DM has been studied before in the context of DM capture in the Sun and the solar system, with a focus on potential indirect DM signals~\cite{Damour:1998rh,Damour:1998vg,Lundberg:2004dn,Peter:2009mi,Peter:2009mm}. On the contrary, here we study the DM capture by the Earth, and investigate the as of yet unexplored consequences for DM direct detection.

\section{DM capture by the Earth}
The capture of DM particles by stellar objects and the Earth has been studied extensively in the past~\cite{Press:1985ug,Gould:1987ir,Kouvaris:2007ay,Catena:2015uha}.  In particular, the capture of DM in the Sun and its subsequent distribution in bound elliptical orbits has been studied both analytically~\cite{Damour:1998rh,Damour:1998vg} and numerically~\cite{Peter:2009mi,Peter:2009mm}. Here we focus on DM capture by the Earth. The key point for the DM capture is that the particle should scatter underground to velocities that are below the escape velocity of that particular point of the Earth, thus leading to a gravitational bound orbit. 
The rate of scattering events (within the infinitesimal volume $d^3 {\bf x}$) to velocities below the escape velocity that lead to elliptical orbits of angular momentum $J$ within $[J,^2 J^2+dJ^2]$ is given by~\cite{Damour:1998vg} 
\begin{align}
\label{dNdot}
 d\dot{N}_{A} = \frac{2\pi \sigma_Av f(v)n_A(r) }{J_{\rm max}^2 \beta_{+}^A} &\left(1-\frac{J^2}{J^2_{\rm max}}\right)^{-1/2} F_A^2(Q)\nonumber \\
&\times  \Theta_\alpha \Theta_{J} \left( d^3{\bf x}\,dv\right)\, d\alpha \, dJ^2 \,.
\end{align}
 $\alpha=G M_{\oplus}/a$, where $a$ is the semi-major axis of the elliptical orbit the DM particle scatters to,  $G$ is the Newton constant and $M_{\oplus}$ is the Earth's mass.  $f(v)$ is the DM velocity distribution before scattering.
Eq.~(\ref{dNdot}) assumes scattering by a single element of mass number $A$ and number density $n_A(r)$ at a distance $r$ from the Earth's centre.~The DM-nucleus scattering cross-section at zero momentum transfer $\sigma_A$ can be written as 
$\sigma_A =  \sigma_{n} A^2 \mu^2_{ A}/\mu^2_{ n}$,
where $\mu_{ A}$ and $\mu_{n}$ are respectively the DM-nucleus and DM-nucleon reduced masses and $\sigma_{n}$ is the so-called spin-independent (SI) DM-nucleon scattering cross section. $J_{\rm max}=r(v_1^2-\alpha)^{1/2}$ is the maximum angular momentum for a bound orbit at $r$, and 
$\beta_{\pm}^{A}=4 m_\chi m_A/(m_\chi\pm m_A)^2$
with $m_\chi$ and $m_A$ being the DM and nucleus masses respectively. The step functions 
$
\Theta_\alpha\equiv \Theta\left[ \beta_{-}^A\left( v_1^2-\frac{\alpha}{\beta_{+}^{A}} \right)-v^2+v_1^2\right],~
\Theta_J\equiv\Theta(J_{\rm max} - J)
$
guarantee that only kinematically allowed bound orbits are populated. $F_A(Q)$ is the nuclear form factor associated with $\sigma_A$, accounting for the loss of coherence in scattering of energy transfer $Q$~\cite{Damour:1998vg}.~Liouville's theorem dictates that the velocity distribution remains constant along the trajectory of a particle i.e. $f(r,v)=f_{\infty}(v_{\infty})$, where $f(r,v)$ and $f_{\infty}(v_{\infty})$ are the velocity distributions just before the collision and  at asymptotically far away distances from the Earth and are related as $v_{\infty}^2=v^2-v_{\text{esc}}^2(r)$. Upon assuming a Maxwell-Boltzmann distribution for $f_{\infty}(v_{\infty})$, and after averaging over the angle of attack of the DM particle, the velocity distribution just before the collision is
\begin{align}
\label{fffu}
f(v)  dv=\frac{n_\chi}{4\pi^{3/2}v_E v_0\sqrt{v^2-v_1^2}}\left(e^{-\frac{v_{-}^2}{v_0^2}}-e^{-\frac{v_{+}^2}{v_0^2}}\right)  dv\,,
\end{align}
where $v_{\pm}=\sqrt{v^2-v_1^2}\pm v_E$, $v_1$ is the Earth's escape velocity, $v_0=220$~km~s$^{-1}$ is the local standard of rest, $v_E=232$~km~s$^{-1}$ is the Earth velocity in the galactic rest frame, and $n_\chi$ the DM number density in the Earth's neighborhood. The above distribution is in the rest frame of the Earth.
The escape velocity of the Earth varies from 15~km~s$^{-1}$ at the Earth's centre to 11.2~km~s$^{-1}$ at the Earth's surface. Since the variation is small,  we simplified our calculation, by setting the escape velocity to its surface value $v_1=$11.2~km~s$^{-1}$. This makes $f(r,v)$  independent of $r$ (leading to  Eq.~(\ref{fffu})). 

Eq.~(\ref{dNdot}) expresses the rate of trapped particles per unit volume of the Earth in terms of the semi-major axis (via $\alpha$) and angular momentum $J$ of the orbit after the collision. It is however more convenient for the calculation of the bound DM flux on the detectors to express the rate in terms of the ellipticity of the orbit $e$ and pericenter (the closest distance of the orbit to the center of the Earth) $r_m$. In this case Eq.~(\ref{dNdot}) can be rewritten as
\begin{align}
\label{dNdot2}
 d\dot{N}_A= 4\pi G M_{\oplus} \frac{\sigma_A v f(v) n_{A}(r)}{r^2\beta_{+}^A}
\left(1-\frac{r_m^2}{r^2} \right)^{-1/2}F^2_A(Q)  \nonumber \\
\times\Theta_{r_{m}}\Theta_e \left( d^3{\bf x}\,dv\right)\, d e \,dr_m \,.
\end{align}
The condition $J=r_{m}(v_1^2-\alpha)^{1/2}\le J_{\rm max}$ imposed by $\Theta_J$ becomes $\Theta_{r_m}\equiv\Theta(r-r_m)$ and $\Theta_e$ is $\Theta_\alpha$ having subsituted $\alpha=GM_{\oplus}(1-e)/r_m$. Recall that the semi-major axis $a=r_m/(1-e)$.~Eq.~(\ref{dNdot2}) should be summed over all elements abundant in the Earth. In practice we take into account the most abundant elements, i.e. $^{16}$O, $^{28}$Si, $^{24}$Mg, $^{56}$Fe, $^{40}$Ca, $^{23}$Na, $^{32}$S, $^{59}$Ni, and $^{27}$Al assuming the standard composition and density profile of chemical elements in the Earth $n_A(r)$ provided in~\cite{Gondolo:2004sc}. Integrating Eq.~(\ref{dNdot2}) over $d^3{\bf x}\, dv$ and summing over elements gives
\begin{align}
\label{dNdot3}
& d \dot{N} = 16\pi^2 G M_{\oplus} \sum_A \frac{\sigma_A}{\beta_{+}^A} K_A(r_m,e) \nonumber \\ \times&\int_{r_m}^{R_{\oplus}} d r\, n_A(r) 
\left(1-\frac{r_m^2}{r^2} \right)^{-1/2}  d e \,{\rm d}r_m \equiv g(r_m,e)  de \, dr_m.
\end{align}
Eq.~(\ref{dNdot3}) gives the rate  of  accumulation of trapped DM particles into bound elliptical orbits of ellipticity within $[e,e+{\rm d}e]$, and pericenter within $[r_m,r_m+{\rm d}r_m]$.~In the derivation of Eq.~(\ref{dNdot3}), we have assumed spherical symmetry, i.e. ${\rm d}^3{\bf x}=4\pi r^2 {\rm d}r$. $K_A(r_m,e)$ is defined as
\begin{align}
\label{eq:K}
K_A(r_m,e) \equiv  \int  dv \,v f(v) F^2_A(Q) \Theta_e =\int_{v_1}^{v_2}  dv \,v f(v) F^2_A(Q),
\end{align}
where $Q=(1/2)m_{\chi} (v^2-v_1^2+GM_{\oplus}(1-e)/r_m)$ is the energy loss in the collision. The upper limit $v_2$ comes from the step function $\Theta_e$ and it is given by
$
v_2=\sqrt{(1+\beta_{-}^A) v_1^2 - \frac{G M_\oplus}{r_m}(1-e)\frac{\beta_{-}^A}{\beta_{+}^A}} \,.
$
The lower limit of intergration is obviously the escape velocity $v_1$ since a DM particle with zero speed at asymptotic far distances from the Earth, will acquire $v_1$ once it reaches the Earth. 

\section{Recoil Energy Spectrum of Bound Dark Matter}
In this section we calculate the recoil energy spectrum due to bound DM particles. In order to do this, we must first estimate the time it takes for a particle orbiting the Earth to scatter for a second time. We make an approximation here, i.e. we  consider recoil events in direct detection experiments that are produced by DM particles bound in elliptical orbits that have scattered only once inside the Earth before reaching the detector. Although in principle even particles that have scattered more than once can contribute to the spectrum, we expect that successive collisions will lead to a diminished DM kinetic energy which practically means very low recoil energies. Therefore within this approximation, we estimate the number of DM particles that can accumulate in different orbits and have scattered only once. The number of periods $N$ required for a DM particle to scatter for a second time is 
\begin{align}
\label{NN}
N=\left (\sum_A \int_0^{\phi_1} n_A(r) \sigma_A \xi(r_m,e) d\phi \right)^{-1},
\end{align}
where $\xi(r_m,e)d\phi$ is an infinitesimal  path along the elliptic trajectory of the orbit given by $\xi(r_m,e)= 2 r_m (1+e)\sqrt{1+e^2+2e\cos\phi}/(1+e\cos\phi)^2$. The integration is along the underground part of the orbit. $\cos\phi_1 = \frac{r_m}{R_\oplus} \frac{(1+e)}{e} - \frac{1}{e}$ corresponds to the angle subtended by the pericenter and the point where the orbit crosses the Earth ($r =R_\oplus$) from the Earth's center. The condition $-1<\cos\phi_1<1$ implies that
$
\frac{1-e}{1+e} \le \frac{r_m}{R_\oplus} \le 1\,.
$
For a given orbit, the time $T(r_m,e)$ a DM particle can  spend without scattering for a second time is on average
\begin{align}
\label{eq:Gamma}
T(r_m,e)\equiv \min [ N \times \tau(r_m,e), \tau_\oplus] \,,
\end{align}
where $\tau_\oplus\simeq 4.5\times 10^9$ years is the age of the Earth and
$\tau(r_m,e)= \sqrt{\frac{4\pi^2}{G M_\oplus} \frac{r_m^3}{(1-e)^3}}$ is the period of the elliptical bound orbit.

\begin{figure*}[t]
\begin{minipage}{0.32\textwidth}
\begin{center}
\includegraphics[width=\textwidth]{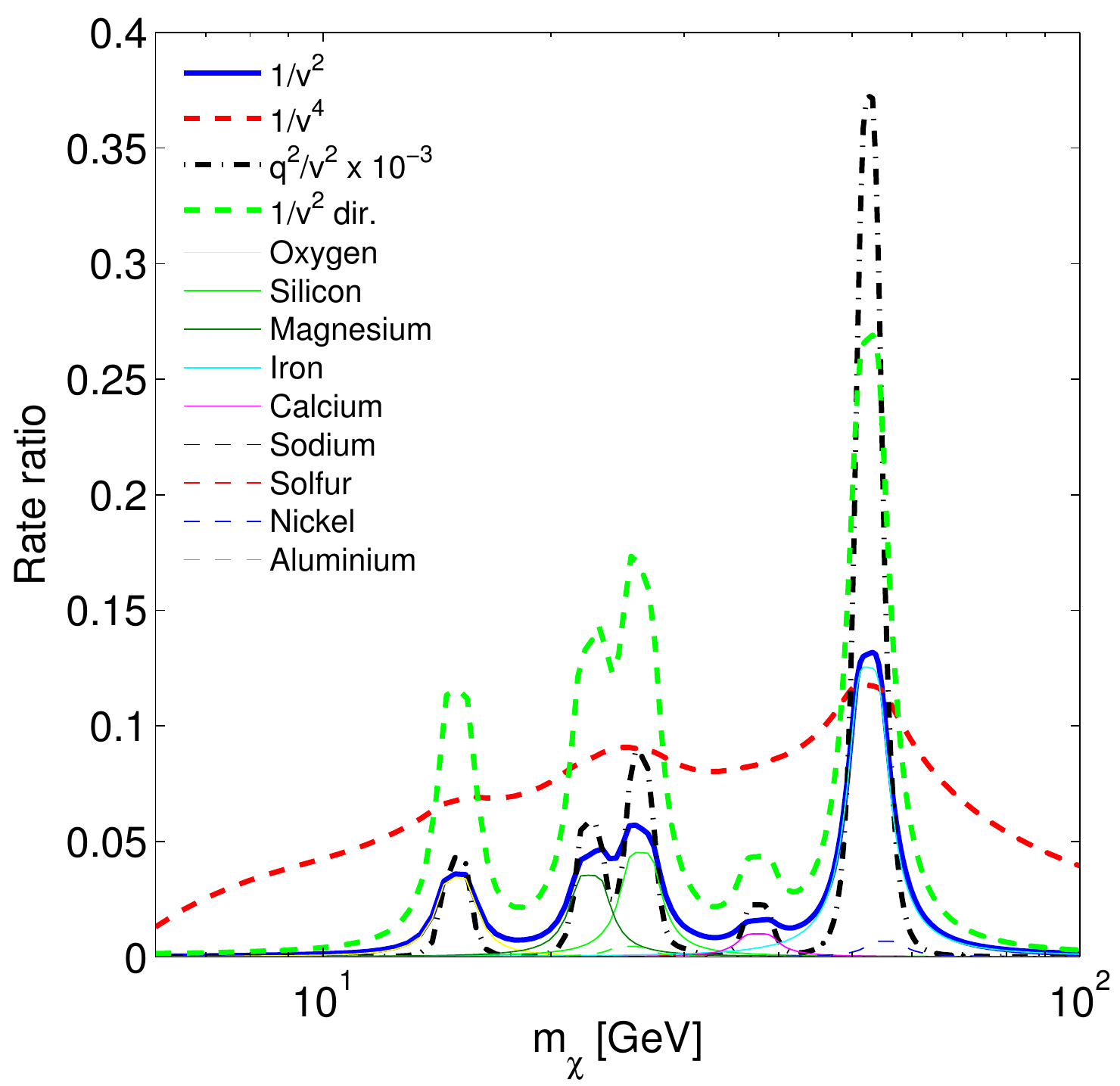}
\end{center}
\end{minipage}
\begin{minipage}{0.32\textwidth}
\begin{center}
\includegraphics[width=\textwidth]{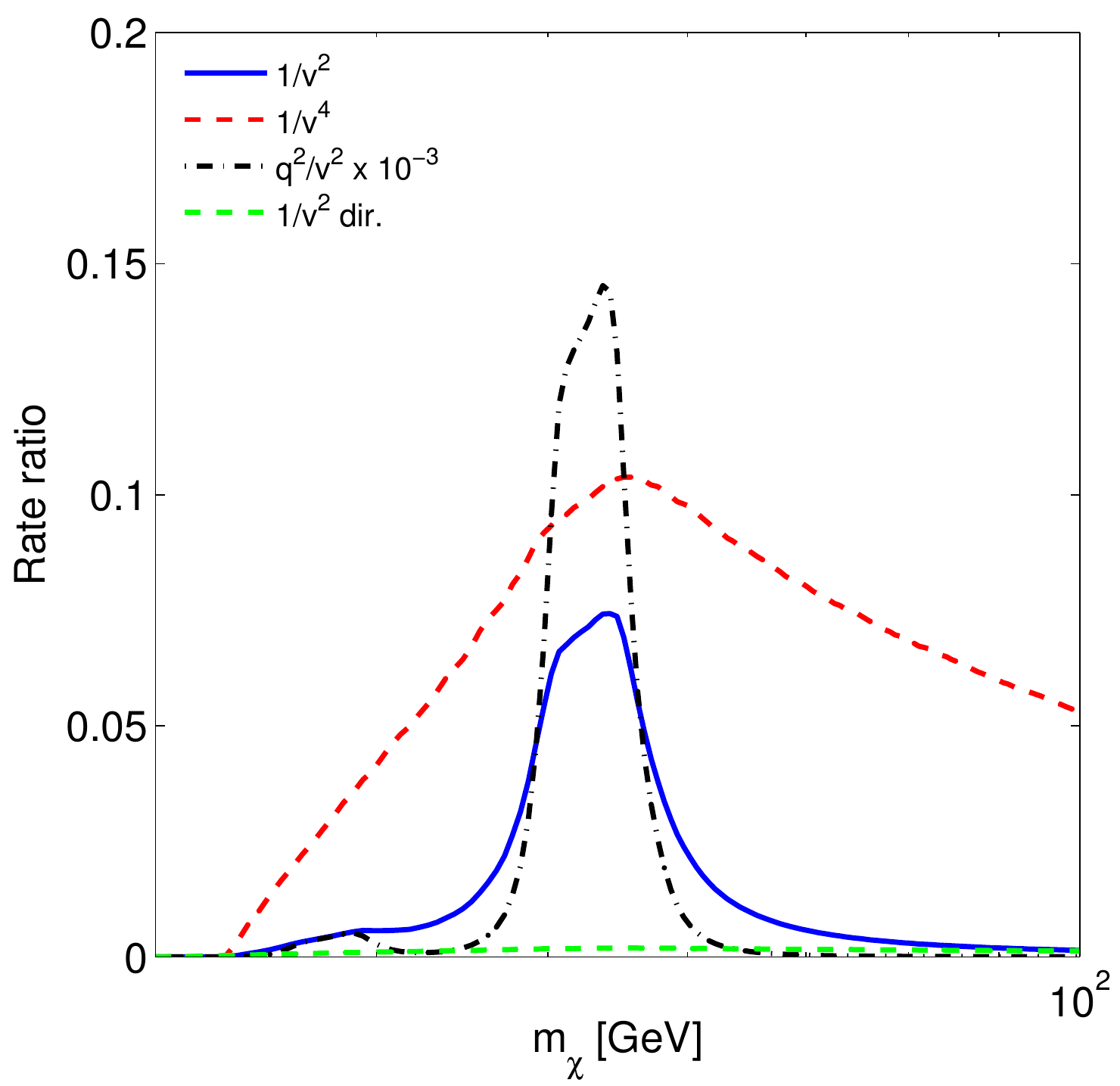}
\end{center}
\end{minipage}
\begin{minipage}{0.32\textwidth}
\begin{center}
\includegraphics[width=\textwidth]{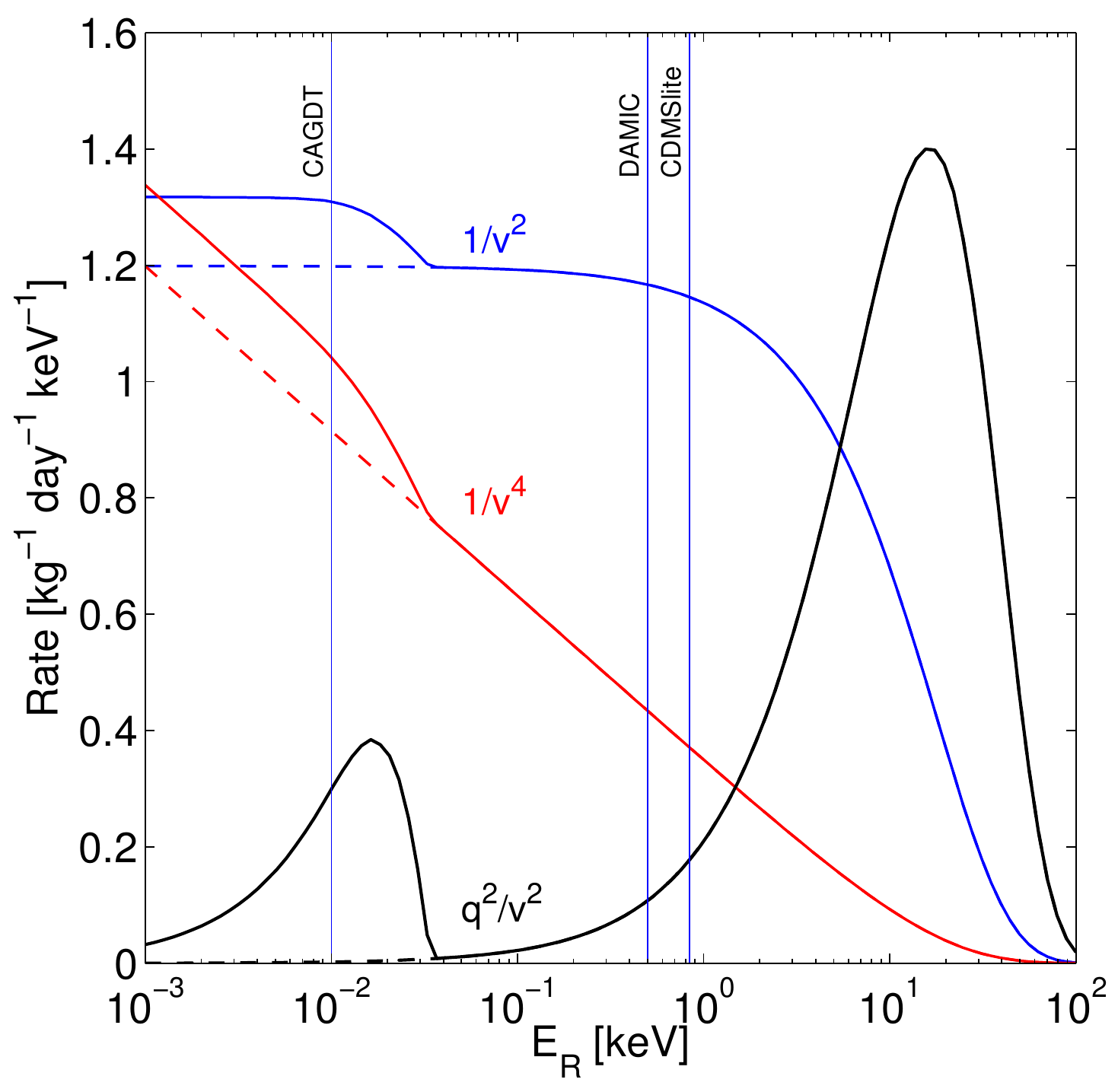}
\end{center}
\end{minipage}
\caption{{\it Left panel.} Ratio of the differential rate of recoil events induced by bound DM over the one induced by ordinary halo DM as a function of $m_\chi$ at recoil energy 1~eV.~We consider a Germanium detector with perfect energy resolution, and three dark matter-nucleon interaction types with differential cross-sections scaling as reported in the legend.~Thin colored lines assume capture by single elements in the Earth and a differential cross-section $\propto 1/v^2$.~For this interaction type we also consider the ratio of directional rates. Note that the ratio for $q^2/v^2$ interactions must be rescaled by a factor of $10^3$.~{\it Central panel.} Same as for the left panel for a recoil energy of 20~eV.~{\it Right panel.} Arbitrarily normalized differential rates of recoil events as a function of the recoil energy for a DM particle mass of 50~GeV.~Solid lines correspond to total rates, while dashed lines to the associated halo DM particle contribution.~Vertical lines show the threshold energies of selected present and projected experiments (for reference, we also include DAMIC, though it uses Silicon as target material).}
\label{fig:plots}
\end{figure*}

The differential event rate for a given orbit characterized by $r_m$ and $e$ is
\begin{align}
\frac{dR_{r_m,e}}{dE_R}=N_T \frac{d\sigma}{dE_R} \mathcal{F}=N_T \frac{d\sigma}{dE_R}\frac{{\rm d} \dot{N}}{4 \pi l_c^2}\frac{2T(r_m,e)}{\tau (r_m,e)}, \label{difrate}
\end{align}
where $N_T$ is the number of target nuclei in the detector. $\mathcal{F}$ is the flux of bound DM particles in orbits of pericenter $r_m$ and ellipticity $e$ crossing the detector. $d \dot{N}$ is the rate with which a particular orbit is populated (see Eq.~(\ref{dNdot3})). This orbit can accumulate DM particles for a time interval $T(r_m,e)$. During each revolution of period $\tau (r_m, e)$, a DM particle crosses twice the sphere of radius $\ell_c$ which is the location of the detector. We assume that DM crosses the surface of the sphere isotropically (thus the factor $4\pi \ell_c^2$).
For the SI contact interactions 
$
\frac{d \sigma}{dE_R}=\frac{m_N \sigma_n A_N^2F^2(E_R)}{2 \mu_n^2 v^2}, \label{cross}
$
where $A_N$ corresponds to the number of nucleons in the detector-target nucleus $N$. The velocity $v$ of bound DM particle at the detector is completely determined by $r_m$ and $e$ and can be easily shown to be 
\begin{align}
\label{vv}
v=\sqrt{2 G M_{\oplus} \left (\frac{1}{l_c}-\frac{1-e}{2 r_m} \right ) }.
\end{align}

Combining Eqs.~(\ref{dNdot3}), (\ref{eq:Gamma}) and (\ref{difrate}) we obtain the differential rate of events
\begin{align}
\label{eq:rec}
\frac{dR}{dE_R}= \kappa & \int_0^1 \int_{\frac{1-e}{1+e}R_\oplus}^{R_\oplus}de  dr_m  \frac{g(r_m,e)}{v^2}\frac{T(r_m,e)}{\tau(r_m,e)} dr_m de,
\end{align}
where
$
\kappa = N_T m_N \sigma_n A_N^2 F^2(E_R)/(4 \pi \ell_c^2 \mu_n^2)$. 

We also study  the spectrum of  bound DM scattering off directional detectors. In particular we look into the spectrum of recoils coming from a direction perpendicular to the vector that connects the center of the Earth with the detector.  We have found that such horizontal directions give an enhancement in the bound/halo ratio of DM events in the detector. Generically the directional rate for energy recoil $E_R$ and recoil direction within the solid angle $d\Omega_q$ is $\frac{dR}{dE_Rd\Omega_q}=N_T\int\frac{d\sigma}{dE_Rd\Omega_q}d\Phi$, where $d\Phi$ is the flux of particles arriving at the detector. For SI contact interactions with nucleons, the rate of events for bound DM is
\begin{align}
\label{direct1}
\frac{dR}{dE_R d \Omega_q}= &\frac{\kappa_d}{\delta \ell_c^2} \int \frac{1}{v^2} \delta \left ( \cos\theta_q-\frac{v_{\text{min}}}{v} \right ) \frac{g(r_m,e)}{\tau(r_m,e)} T(r_m,e) \nonumber \\
&\times dr_m de\frac{d\cos\theta d\phi}{4\pi}\frac{d\omega}{2\pi},
\end{align}
where $\kappa_d=N_T m_N \sigma_n A_N^2F^2(E_R)/(4 \pi \mu_n^2)$, $v_{\text{min}}$ is the minimum DM velocity that can induce a nuclear recoil of energy $E_R$, $\delta \ell_c$ is the characteristic length of the detector (which eventually will drop out), $g(r_m,e)$ is defined in Eq.~(\ref{dNdot3}), $\theta$ and $\phi$ are the polar angles of the location of the pericenter $r_m$ of the orbit (with the $z$-axis being the vector from the center of the Earth to the detector) and $\omega$ is the angle that defines the tilt of the elliptical plane of the orbit. The delta function determines the kinematically  allowed recoil angles between the incoming DM particle and the nucleus recoil direction $\theta_q$. For horizontal recoil directions (i.e. perpendicular to the direction of gravity at the detector) $\cos\theta_q=\pm (1+e\cos\theta ) \cos\phi/\sqrt{1+e^2+2 e \cos\theta}$, where $\pm$ corresponds to the two possible directions of motion i.e. (counter)clockwise. For the orbits to pass through the volume $\sim \delta \ell_c^3$ of the detector, $\delta r_m =\delta \ell_c(1+e \cos\theta)/(1+e)$ and $\delta \omega =\delta \ell_c / ( \ell_c \sin \theta)$. By using the above and performing the integration over $\theta$ with the use of the delta function, Eq.~(\ref{direct1}) reads
 \begin{align}
\label{direct2}
\frac{dR}{dE_R d \Omega_q}= &\frac{\kappa_d}{2 \pi^2 \ell_c} \int_{e_1}^1 de \int_{\phi_a}^{\phi_b} d\phi \frac{1}{v^2}\frac{g(r_m,e)}{\tau(r_m,e)} T(r_m,e)  \nonumber \\ &\times \frac{1+e y}{1+e} \frac{1}{\sqrt{1-y^2}}
 \frac{\sqrt{\frac{2 v_{\text{min}}^2\ell_c}{G M_{\oplus}}-(1-e^2)\cos^2\phi}}{e \cos^2\phi},
\end{align}
where $v$ is given by Eq.~(\ref{vv}), $r_m=\ell_c (1+e y)/(1+e)$, $y= (-\cos^2\phi+\gamma)/(e\cos^2\phi)$ and $\gamma=v_{\text{min}}^2\ell_c/(G M_{\oplus})$. The integration limits are $e_1=\text{Max}[\gamma-1,0]$, $\phi_a=\cos^{-1} (\text{Min}[1,\sqrt{\gamma/(1-e)}])$ and $\phi_b=\cos^{-1}\sqrt{\gamma/(1+e)}$.~In the numerical applications we set $l_c=R_{\oplus}-2$~km where $R_{\oplus}$ is the radius of the Earth and 2 km is the depth of the detector.
\section{Results}
 In Fig.~\ref{fig:plots} we show our main results.~The left (central) panel shows the ratio of events per recoil energy induced by bound DM over ordinary halo DM evaluated at a recoil of 1 (20) eV for different types of interactions as a function of the DM mass. Note that although the plots have been made  assuming a cross section of  $10^{-45}$~cm$^2$ (for the familiar SI interaction), we have found the ratio to be almost independent of the DM-nucleus cross section. This is easy to explain. The rate of bound DM events in the detectors has three entries where the DM-nuclei cross section plays a role, i.e. in the DM scattering in the detector, in the capture rate and in the time a specific orbit can accumulate DM without rescattering. The first dependence is common for both halo and bound DM events. However the other two entries basically cancel each other out. This is because larger DM-nuclei cross sections will lead to higher capture rates but lower accumulation times since the probability for rescattering is higher. We have plotted the ratios for three different interactions where $d\sigma/dE_R$ scales as $1/v^2$ (the typical contact SI interaction, for which we also consider the ratio of directional rates assuming Fluorine as target material), $1/v^4$ and $q^2/v^2$ which is that of the operator $O_{11}$ (or $O_{9}$, $O_{10}$, and $O_{12}$) in, e.g.~\cite{Fitzpatrick:2012ix,Catena:2015uha} (see also~\cite{Catena:2015vpa,Kavanagh:2015jma} for a discussion on effective operators in the context of directional detectors). Although Eq.~(\ref{eq:rec}) has been derived for the first type (contact SI), it is easy to derive the analogous expressions for the other two following the same steps leading to Eq.~(\ref{eq:rec}), keeping in mind that the different $v$ and $q$ dependence of $d\sigma/dE_R$ will affect $K_A(r_m,e)$ of Eq.~(\ref{eq:K}),  $\sigma_A$ in Eq.~(\ref{NN}) and the final recoil that takes place in the detector.

One can see that each element in the Earth produces a characteristic resonance in the differential rate of bound DM, the most pronounced of which is associated with Iron. Note also that the resonance effect reduces and smooths out as one goes to higher recoil energies (e.g. 20 eV in the central panel). Bound DM can give an increase of the order of $10\%$ for $1/v^2$ and $1/v^4$  (at 1 eV).~For horizontal recoils in directional detectors we find that the bound DM rate is up to $25\%$ of the halo one. This would practically mean that for every 4 halo DM events there will be one from bound DM. The result is more dramatic for interactions with momentum dependence like the operator $O_{11}$, where (as it can be seen) the bound DM rate can be up to $\sim 350$ times larger than the halo one. Notice that  in Fig.~\ref{fig:plots} for illustrative purposes, the ratio associated with $O_{11}$ has been multiplied by $10^{-3}$. 

The right panel in Fig.~\ref{fig:plots} shows the rate of events  as a function of the recoil energy for a fixed DM mass of 50 GeV. We have included current (and projected) energy thresholds of various experiments as a reference~\cite{Agnese:2013jaa,Chavarria:2014ika, CAGDT}. For the case of contact SI and $1/v^4$ interactions, the bound DM contribution  appears in the spectrum  as a bump at low energies. The height of the bump with respect to the ordinary halo spectrum is given by the ratio plotted in the previous figures we discussed.~For the momentum dependent interaction $O_{11}$, the spectrum appears with two distinct peaks, one at high energy  - where halo DM dominates - and one at low energy, where bound DM is the dominant source (up to several hundreds larger than the halo DM, as we mentioned earlier).~With a relative ratio on the height of the two peaks halo/bound=1.4/0.4=3.5, for every 3.5 high energy events there will be one (unexpected) low energy event due to scattering of bound DM on the detector. 

In this paper we demonstrated the existence of new spectral features in the low recoil energy spectrum of direct DM detectors due to scattering of DM particles that are gravitationally bound to the Earth. These features are universal i.e.~the ratio of bound DM events over halo ones is independent of the DM-nucleus cross section. This could be quite useful in distinguishing beyond doubt a DM signal from possible background and it could identify the specific velocity and momentum dependence of the DM-nucleon interaction, since different types of interactions create qualitatively different spectral features. Since the low energy part of the recoil spectrum contains the key information described here, the effort for lowering the experimental energy thresholds is of great importance, as  it would allow to reveal new features that can identify DM beyond any doubt.

{{\it Acknowledgments.}~CK is partially funded by the Danish National Research Foundation, grant number DNRF90.


\begin{thebibliography}{99}



\bibitem{Damour:1998rh} 
  T.~Damour and L.~M.~Krauss,
  Phys.\ Rev.\ Lett.\  {\bf 81}, 5726 (1998)
  doi:10.1103/PhysRevLett.81.5726
  [astro-ph/9806165].


\bibitem{Damour:1998vg} 
  T.~Damour and L.~M.~Krauss,
  Phys.\ Rev.\ D {\bf 59}, 063509 (1999)
  doi:10.1103/PhysRevD.59.063509
  [astro-ph/9807099].

\bibitem{Lundberg:2004dn}
  J.~Lundberg and J.~Edsjo,
  Phys.\ Rev.\ D {\bf 69} (2004) 123505
  doi:10.1103/PhysRevD.69.123505
  [astro-ph/0401113].

\bibitem{Peter:2009mi} 
  A.~H.~G.~Peter,
  Phys.\ Rev.\ D {\bf 79}, 103531 (2009)
  doi:10.1103/PhysRevD.79.103531
  [arXiv:0902.1344 [astro-ph.HE]].


\bibitem{Peter:2009mm} 
  A.~H.~G.~Peter,
  Phys.\ Rev.\ D {\bf 79}, 103533 (2009)
  doi:10.1103/PhysRevD.79.103533
  [arXiv:0902.1348 [astro-ph.HE]].


\bibitem{Press:1985ug} 
  W.~H.~Press and D.~N.~Spergel,
  Astrophys.\ J.\  {\bf 296}, 679 (1985).
  doi:10.1086/163485


\bibitem{Gould:1987ir} 
  A.~Gould,
  Astrophys.\ J.\  {\bf 321}, 571 (1987).
  doi:10.1086/165653


\bibitem{Kouvaris:2007ay} 
  C.~Kouvaris,
  Phys.\ Rev.\ D {\bf 77}, 023006 (2008)
  doi:10.1103/PhysRevD.77.023006
  [arXiv:0708.2362 [astro-ph]].

\bibitem{Catena:2015uha} 
  R.~Catena and B.~Schwabe,
  JCAP {\bf 1504}, no. 04, 042 (2015)
  doi:10.1088/1475-7516/2015/04/042
  [arXiv:1501.03729 [hep-ph]].



\bibitem{Fitzpatrick:2012ix}
  A.~L.~Fitzpatrick, W.~Haxton, E.~Katz, N.~Lubbers and Y.~Xu,
  JCAP {\bf 1302} (2013) 004
  doi:10.1088/1475-7516/2013/02/004
  [arXiv:1203.3542 [hep-ph]].

\bibitem{Gondolo:2004sc}
  P.~Gondolo, J.~Edsjo, P.~Ullio, L.~Bergstrom, M.~Schelke and E.~A.~Baltz,
  JCAP {\bf 0407} (2004) 008
  doi:10.1088/1475-7516/2004/07/008
  [astro-ph/0406204].

\bibitem{Catena:2015vpa}
  R.~Catena,
  JCAP {\bf 1507} (2015) 07,  026
  doi:10.1088/1475-7516/2015/07/026
  [arXiv:1505.06441 [hep-ph]].

\bibitem{Kavanagh:2015jma}
  B.~J.~Kavanagh,
  Phys.\ Rev.\ D {\bf 92} (2015) 2,  023513
  doi:10.1103/PhysRevD.92.023513
  [arXiv:1505.07406 [hep-ph]].

\bibitem{Agnese:2013jaa}
  R.~Agnese {\it et al.} [SuperCDMS Collaboration],
  Phys.\ Rev.\ Lett.\  {\bf 112} (2014) 4,  041302
  doi:10.1103/PhysRevLett.112.041302
  [arXiv:1309.3259 [physics.ins-det]].
  
\bibitem{Chavarria:2014ika}
  A.~E.~Chavarria {\it et al.},
  Phys.\ Procedia {\bf 61} (2015) 21
  doi:10.1016/j.phpro.2014.12.006
  [arXiv:1407.0347 [physics.ins-det]].
  
  \bibitem{CAGDT}
  D.~Mei, talk given at the ``Berkeley Workshop on Dark Matter Detection'', 8-9 June 2015.
  
  

\end{thebibliography}
\end{document}